\documentclass[%
 reprint,
superscriptaddress,
 amsmath,amssymb,
 aps,
]{revtex4-1}

\usepackage{graphicx}
\usepackage[space]{grffile}
\usepackage{latexsym}
\usepackage{textcomp}
\usepackage{longtable}
\usepackage{multirow,booktabs}
\usepackage{amsfonts,amsmath,amssymb}
\usepackage{natbib}
\usepackage{url}
\usepackage{hyperref}
\hypersetup{colorlinks=false,pdfborder={0 0 0}}
\newif\iflatexml\latexmlfalse

\usepackage[utf8]{inputenc}
\usepackage[english]{babel}

\begin{document}

\title{Self-interaction effects on charge-transfer collisions}

\author{Edwin E. Quashie}
\affiliation{Department of Physics, Florida A\&M University, Tallahassee, FL 32307, USA}
\affiliation{Quantum Simulations Group, Lawrence Livermore National Laboratory, Livermore, California 94551, USA}

\author{Bidhan C. Saha}
\affiliation{Department of Physics, Florida A\&M University, Tallahassee, FL 32307, USA}

\author{Xavier Andrade}
\affiliation{Quantum Simulations Group, Lawrence Livermore National Laboratory, Livermore, California 94551, USA}

\author{Alfredo A. Correa}
\affiliation{Quantum Simulations Group, Lawrence Livermore National Laboratory, Livermore, California 94551, USA}

\begin{abstract}
In this article, we investigate the role of the self-interaction error in the simulation of collisions using time-dependent density functional theory (TDDFT) and Ehrenfest dynamics. 
We compare many different approximations of the exchange and correlation potential, using as a test system the collision of $\mathrm{H^+ + CH_4}$ at $30~\mathrm{eV}$. 
We find that semi-local approximations, like PBE, and even hybrid functionals, like B3LYP, produce qualitatively incorrect predictions for the scattering of the proton. 
This discrepancy appears because the self-interaction error allows the electrons to jump too easily to the proton, leading to radically different forces with respect to the non-self-interacting case. 
From our results, we conclude that using a functional that is self-interaction free is essential to properly describe charge-transfer collisions between ions and molecules in TDDFT.%
\end{abstract}%

\maketitle

\section{Introduction}

When a slow projectile collides with a molecular target, several microscopic processes, such as ro-vibrational excitations, dissociation, charge transfer, nuclear exchange and fragmentation occur~\cite{Udseth_1973,Barnett_1977}. 
These processes have received considerable attention in a wide range of applied fields, such as radiation therapy, material science, accelerators, plasma physics, chemistry and astrophysics~\cite{McDowell_1980,McCarroll_1982,Salehzadeh_2015}. 
At low incident energies, a large number of ion-molecule collisions experiments have been reported~\cite{Gentry_1975,Noll_1986,Niedner_1987,Aristov_1991}; full understanding of these collision processes still poses challenges to both experiment and theory~\cite{Krutein_1979,Friedrich_1987,Chiu_1988,Aristov_1993}. 

The development of practical methods for coupled molecular dynamics and real-time electron dynamics, have provided new effective tools for studying the atomic and molecular collisions with the predictive power of first principles methods~\cite{Saalmann_1996,Yabana_1998,Nagano_1999,Nagano_2000,Tong_2002,Henkel_2009,Wang_2011,Avenda_o_Franco_2012,Hong_2016,Quashie_2016}. 
Ehrenfest dynamics~\cite{L_dde_2003,Isborn_2007,Castro_2013} combined with time-dependent density functional theory~(TDDFT)~\cite{Runge_1984} is one of the prominent method to understand collision processes.

As in all practical first-principles simulation methods, approximations are required to reduce the complexity of the full Schr\"ondiger equation. 
The two major approximations in Ehrenfest-TDDFT are to approximate the dynamics of the ions as a single classical trajectory, and the use of an approximated exchange and correlation term in the TDDFT electron equations. 
In this paper we will focus on this second approximation, by comparing the results due to various functionals for the same process. 
While the findings of the TDDFT tend to get qualitatively good agreement with experimental results even with the most basic approximations to the exchange and correlation functional~\cite{Engel_2010,Tsuneda_2014}, we find that for collision processes it becomes essential to use more advanced functionals that avoid the self-interaction error. 
Since the development of the first practical DFT approaches, self-interaction effects have been recognized as the prime sources of errors for many applications~\cite{Heaton_1983,Johnson_1994,Zhang_1998,POLO_2002,Lundberg_2005,Ruiz_2005,Mori_Sa_nchez_2006,Pederson_2014}.

The dynamics of ion-molecule interactions at low collision energies has enjoyed wide applications especially for the ability of the incoming ion to capture electrons from the target. 
As a model system for our study we choose the collision of a proton $(\mathrm{H^+})$ with a methane molecule, which is a prototypical example at the intersection between the fields of chemistry and atomic collisions~\cite{Bransden_1993}.

The process $\mathrm{H^+ + CH_4}$ has been studied for $10 \leq E \leq 50~\mathrm{eV}$ by several groups. 
Toennies \emph{et al.} \cite{Niedner_1987,Friedrich_1987,Chiu_1988}, Udseth \emph{et al.}~\cite{Udseth_1974}, and Linder and Krutein~\cite{Krutein_1979} used a crossed-beam experiment to study inelastic and charge-transfer processes involving ion and molecule collisions. 
Jacquemin \emph{et al.} \cite{Jacquemin_1997} calculated the differential cross section and integral cross sections and energy loss spectra for the process $\mathrm{H^+ + CH_4}$ at $30~\mathrm{eV}$ using an Electron-Nuclear Dynamics (END) method. 
Their differential cross sections agree nicely with the experimental results for the non-transfer processes. 
Gao \emph{et al.} \cite{Gao_2014} investigated the dynamical evolution related to self-interaction correction (SIC) and calculated fragment intensity and intra-molecule energy transfer for $\mathrm{H^+ + CH_4}$ at $30~\mathrm{eV}$.
%
%

\section{Theory}

A fully quantum simulation of a collision process is not computationally feasible, so some approximations are essential. 
The most fundamental approximation is the separation of the electronic and nuclear degrees of freedom, which is justified by the large difference in mass between these two types of particles. 
Here we use the Ehrenfest approach~\cite{Ehrenfest_1927}, where the dynamics of the electrons are quantum, while the nuclei follow a classical Newtonian dynamics with forces given by the gradient of the instantaneous energy. 
The advantage of this approach is that the dynamics of the electrons are explicitly modeled, so that the simulation includes the excitation of the electrons due to the motion of the ions during the collision.

Even without the ionic degrees of freedom at play, simulating the full quantum dynamics of the electrons is rather impractical, so it requires further approximations. 
An efficient approach to do this simulations is the TDDFT framework~\cite{Runge_1984,Gross_1995,Ullrich_2011,Harker_2013}, where the many-body problem is mapped to the propagation of a non-interacting system.
 
The Ehrenfest-TDDFT equations are (atomic units are used)
\begin{equation}
M_J\frac{\mathrm{d}^2 \vec{R}_J}{\mathrm{d} t^2}=-\frac{\partial E[n, \{\mathbf{R}_J(t)\}_J]}{\partial \vec{R}_J},
\end{equation}
%
\begin{equation}
\mathrm{i}\frac{\partial}{\partial t}\varphi_{i}(\mathbf{r}, t) = \left\{-\frac{\nabla^2}{2} + V_\text{KS}[n, \{\mathbf{R}_J(t)\}_J](\mathbf{r}, t)\right\}\varphi_i(\mathbf{r}, t),
 \label{eq:tdks1} 
\end{equation}
where the electronic density is given by 
\begin{equation}
n(\textbf r, t) = \sum_{i}{|\varphi_i(\textbf r, t)|}^2.
\label{eq:tdks2}  
\end{equation}
$\mathbf{R}_J(t)$ denotes the atomic coordinates evolving according to Ehrenfest forces~\cite{Andrade_2009}, 
and $\varphi_i$ are the electronic orbitals.
The Kohn-Sham (KS) effective potential, defined as $V_\text{KS}[n](\mathrm{r}, t) = \frac{\delta E}{\delta n(\mathbf{r},t)}$ is conceptually partitioned in three contributions:
\begin{multline}
V_\text{KS}[n, \{\mathbf{R}_J\}_J](\textbf r, t) = V_\text{ext}[\{\mathbf{R}_J\}_J](\mathbf{r}, t) \\ 
+ V_\text{H}[n](\mathbf{r}, t) + V_\text{xc}[n](\mathbf{r}, t)
\label{eq:tdks3}
\end{multline}
where $V_\text{ext}(\mathbf{r}, t)$ is the external potential due to ionic core potential (moving ions during the collision) and other external perturbations (not present in this problem). 
$V_\text{H}[n](\mathbf{r}, t)$ is the Hartree potential which describes the classical electrostatic interactions between electrons, and $V_\text{xc}[n](\mathbf{r}, t)$ denotes the exchange-correlation (XC) potential.

The exact form of the XC potential as a functional of the time-dependent density is rather unknown in TDDFT, so this is the part that requires much attention and developing suitable approximation. 
Thus the development of accurate and computationally efficient XC approximations remains a challenge in TDDFT, as this is one of the main sources of error in the theory~\cite{Gould_2013,Liu_2014,Otero_de_la_Roza_2016}.

The self-interaction error is one of the most well known deficiencies of the approximated functionals in both DFT and TDDFT~\cite{Perdew_1981,Mori_Sa_nchez_2006}.
It appears when an electron effectively interacts with itself due to approximations in the XC functional.
The self-interaction error can appear both in the energy as well as in the potential.
We focus on the latter one, as it is the approximations in the XC potential what mainly determines the error in TDDFT simulations.
While it is hard to define the self-interaction error in general, there are simple conditions that a potential must satisfy~\cite{Mori_Sa_nchez_2006}.
The first one is known as the one electron self-interaction condition: for a one-electron system the XC potential must exactly cancel the Hartree potential.
Additionally, to avoid self-interaction in the Coulomb interaction, for an $N$-electron systems each electron must see an effective potential that corresponds to $N-1$ electrons, otherwise each electron is effectively interacting electrostatically with itself. 
As the Hartree potential is the potential of $N$ particles, to compensate, the exchange potential must behave like the potential of a hole and should decay as $-1/r$~\cite{Almbladh_1985}.

Many popular approximated density functionals, including the local density approximation (LDA) \cite{Kohn_1965} and most generalized gradient approximations (GGAs), produce an XC potential that decays exponentially and does not correct for the self-interaction present in the Hatree potential. 
The self-interaction error leads to systems that are too polarizable, as the effective KS potential is less attractive than it should be. 
This induces errors not only in the prediction of polarizabilities and hyperpolarizabilities \cite{Andrade_2007}, but also in ionization potentials~\cite{Merkle_1992,Watanabe_1957} 
and excitation energies \cite{Van_Caillie_2000,Levy_1995}.

In order to 
overcome these errors,
the first alternative is to use an exchange approximation that has the proper asymptotic limit. 
The exact exchange approach (EXX)~\cite{Della_Sala_2002,Wijewardane_2008} 
is such a functional. 
It is, however, quite expensive computationally, especially if one wants to do non-adiabatic molecular dynamics. 
We can use the Krieger-Li-Iafrate (KLI) approximation \cite{Krieger_1990} to reduce the computational cost, 
while retaining the correct asymptotic limit. 
As this is an exchange functional, a properly derived accompanying correlation functional is not yet available.

It is also possible to find some semi-local functionals that have the proper asymptotic limit, in particular GGAs~\cite{van_Leeuwen_1994} and meta-GGAs \cite{Becke_2005,Ra_sa_nen_2010}. 
Hybrid functionals~\cite{Becke_1993} still contain some level of self-interaction since they include only part of Hartree-Fock (HF) exchange (which is self-interaction free). 
To remove it completely, one must use more sophisticated range-separated hybrid functionals that include 100\% exact exchange at long ranges~\cite{Iikura_2001}.

There are several schemes that, instead of proposing a new functional, provide a way to modify an existing approximation in order to remove self interaction~\cite{Perdew_1981,Andrade_2011}. 
Among them, the simplest approach is known as Fermi-Amaldi (FA)~\cite{my_FermiAmaldi}, and consists in simply scaling the Hartree potential by a factor $(N-1)/N$. 
Although it corrects the asymptotic limit, it does not provide accurate exchange potential elsewhere. 
An advanced version of this idea is the average-density self-interaction correction (ADSIC)~\cite{K_mmel_2008,Legrand_2002} which includes additional terms to compensate the errors in the XC potential. 
Both FA and ADSIC have a size-consistency problem which becomes important for collisions; if a system is composed of independent fragments they would provide a correct asymptotic limit for the whole system instead of each fragment.

In order to study the effect of the XC potential in collisions, we sample the results of different approximations and show that they produce qualitatively different results. 
First, we use the standard Perdew-Burke-Ernzerhof (PBE)~\cite{Perdew_1996}, and B3LYP~\cite{Stephens_1994} approximations that contain self-interaction. 
Second, we consider LDA with the self-interaction corrections (SIC) of FA and ADSIC as mentioned earlier.
Finally, we include the exact exchange in the KLI approximation (EXX-KLI). 
We also tested EXX-KLI with the correlation functional from PBE and Lee-Yang-Parr (LYP)~\cite{Lee_1988}, however, we found no significant difference in the results with respect to EXX alone.

\section{Computational Details}

To evaluate the above time-dependent KS (TDKS) equation (Eq.~\ref{eq:tdks1})~\cite{Ullrich_2011a,2006} numerically, finite discretization of the functions $\varphi_i$ as implemented in two different code packages called \textsc{Qbox}~\cite{Gygi_2008} and \textsc{Octopus}~\cite{Castro_2006,Andrade_2015}, were used for the numerical simulations.

In \textsc{Qbox} (with custom time-dependent modifications~\cite{Schleife_2012,Draeger_2016}), a supercell approach is adopted with periodic boundary conditions and the wave functions are expanded in plane waves basis sets. 
In \textsc{Octopus}, however, the TDKS equations are solved by discretizing all quantities in real space~\cite{Rubio_1996,Chelikowsky_1994,Vasiliev_1999}.
These are two \emph{independent} implementations of the same theory, which allows us to rule out effects due to computational inaccuracies and software bugs (specially in the presence of noisy results).

All relevant information regarding the detailed numerical implementations can be found in Refs.~\cite{Schleife_2012,Marques_2003,Gygi_2008,Castro_2004,Andrade_2012,Andrade_2013}. 

Thus, only a brief discussion of both procedures is presented here. 
In our present calculation various XC approximations are used. 
The accuracy of these approximations, however, becomes very important for effective description of various collision dynamics. We have checked that the trends observed with respect to the dependence 
of the functionals are common to all orientations. 

\begin{figure}[h!]
\begin{center}
\includegraphics[width=0.42\columnwidth]{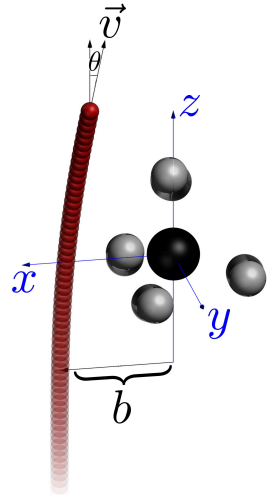}
\caption{{\label{fig:dia_ch4} (color online). 
A schematic diagram for  $\mathrm{H^+ + CH_4}$ initial geometry. 
The red, gray, and black balls represent proton, hydrogen, and carbon respectively. 
The center of mass of the target is in the origin (on the carbon). 
$\theta$ denotes the scattering angle and $b$, the impact parameter.%
}}
\end{center}
\end{figure}

Fig.~\ref{fig:dia_ch4} displays a schematic representation of the initial collisional geometry. 
The nuclei of the ground state $\mathrm{CH_4}$ molecule consists of a carbon and four hydrogen atoms.
There are, however, few different incident orientations of this system. 
Only one orientation as depicted in Fig. \ref{fig:dia_ch4} is used in this calculations; it represents the initial orientation of $\mathrm{H^+ + CH_4}$ before the collision process. 
A single orientation, rather than an integrated result over orientations, facilitates the discussion of the aim of this paper;
the trends observed with respect to the dependency of the functionals are common to all orientations.

This orientation is identical to the Face II orientations of Jacquemin \emph{et al.} \cite{Jacquemin_1997} and Gao \emph{et al.} \cite{Gao_2014} in which 
the incoming proton moves towards the methane from the negative-to-positive z-axis. 
The impact parameter $b$ was increased in the positive x-axis. 
The $\mathrm{C}$ atom of methane is placed at the origin of the coordinates. 
The $\mathrm{CH_4}$ molecule is initially at rest (laboratory reference). 
In order to avoid any prior interactions between the projectile and the methane we initially placed the $\mathrm{H^+}$ at $(b, 0, -10~a_0)$ from the $\mathrm{CH_4}$. 
Initially, the $\mathrm{CH_4}$ is in its electronic ground state, calculated without the presence of the ion. 
The impact parameter \textit{b} is varied in the range of ($0-7.0~a_0$) in steps of $\Delta b = 0.1~a_0$.
The incoming proton initially is given a velocity of $0.0346
~\mathrm{a.u.}$ which corresponds to the kinetic energy $E = 30~\mathrm{eV}$. 
The total simulation time is $19.5~\mathrm{fs}$ with a time step of $\Delta t = 0.838
~\text{attoseconds}$.
In Octopus, we use a spherical simulation domain of radius $20~a_0$ and a grid spacing of $0.3~a_0$.
In \textsc{Qbox}, the simulation domain is a cube of side $30~a_0$ and the plane waves basis set has a $100~\mathrm{Ry}$ energy cutoff.

\section{Results and Discussion}

\begin{figure}[h!]
\begin{center}
\includegraphics[width=1.00\columnwidth]{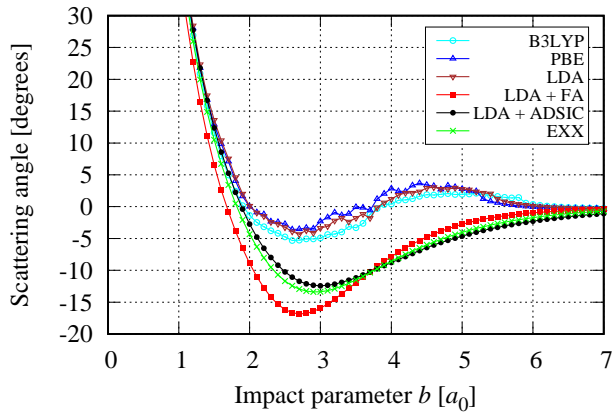}
\caption{{\label{fig:scatt_octopus} (color online). 
Scattering angle $\theta$ as a function of impact parameter $b$ (in units of Bohr radius $a_0$) for different exchange-correlation functionals for Face II orientation (real-space code).
Sign is assigned to the angle according to the quadrant in which the proton scatters (negative angles indicate inwards deflection). 
This is to stress the attractive character of self-interaction and exact exchange theories but partially lost in the B3LYP, PBE and LDA theories.%
}}
\end{center}
\end{figure}

We start our simulation study by calculating the scattering angle $\theta$ as function of impact parameter $b$ for different XC approximations, the results are shown in Fig. \ref{fig:scatt_octopus}, where we observe a significant difference between the outcome of different XC functionals. 
B3LYP, PBE and LDA produced rugged curves, with small scattering angles and two extrema. 
Self-interaction-corrected functionals FA, ADSIC, EXX show a completely different behavior, with a larger negative scattering angle and a single extrema.
If we assume that the interaction with the proton is attractive at long range and repulsive at short range, from molecular collision theory we can obtain some ideas of how the scattering angle should behave~\cite{Child_1976}. 
As the quasi-molecule (\(\mathrm{CH_5^+}\)) is stable~\cite{Chiu_1988}, we expect the interaction between \(\mathrm{H^+}\) and \(\mathrm{CH_4}\) to exhibit such behavior.%

For a small impact parameter the proton will scatter against the repulsive part of the potential, bouncing back and producing large scattering angles with a limit of $180$ degrees for $0$ impact parameter. 
This regime is observed in Fig.~\ref{fig:scatt_octopus} for all functionals.

For higher impact parameters, the proton will probe the attractive region of the scattering potential and it will deflect towards the molecule (negative angles). 
With the increase of the impact parameter the interaction between proton and \(\mathrm{CH_4}\) molecule becomes weaker, so we expect the scattering angle to have a maximum in this region. 
The value of this maximum is known as a the \emph{rainbow angle}, which corresponds to the maximum (absolute) deflection angle for the forward scattering.

In Fig.~\ref{fig:scatt_octopus} we see that functionals with self-interaction corrections exhibit the behavior previously described, however B3LYP, PBE and LDA show a second extrema of positive deflection. 
This suggests that a repulsive interaction is taking place at long ranges due to electronic self-interaction effects~\cite{Schmidt_2014,Gritsenko_2016}.%

\begin{figure}[h!]
\begin{center}
\includegraphics[width=1.00\columnwidth]{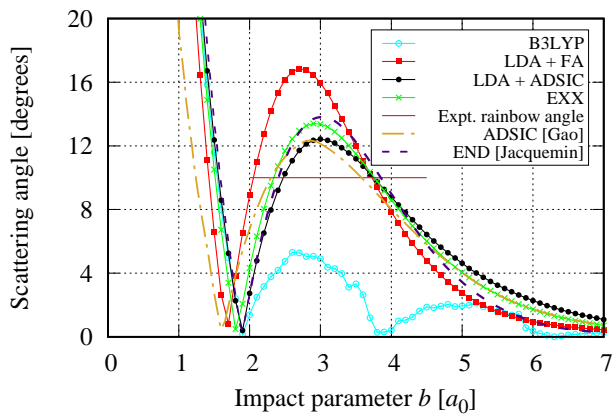}
\caption{{\label{fig:scatt} (color online). 
Scattering angle $\theta$ as a function of impact parameter $b$ (in units of Bohr radius $a_0$) 
for different exchange-correlation functionals compared with theory (Jacquemin \emph{et al.}~\protect\cite{Jacquemin_1997} and Gao \emph{et al.}~\protect\cite{Gao_2014}) for Face II orientation. 
Horizontal line is the experimentally determined rainbow angle over random orientations \protect\cite{Chiu_1988}.%
}}
\end{center}
\end{figure}

In order to further understand the difference in predictions, we compare our results for some of the XC functionals with available theoretical results (see Fig.~\ref{fig:scatt}). 
We also include in the plot the value of the rainbow angle that was measured experimentally by Chiu~\textit{et al.}~\cite{Chiu_1988} (the corresponding impact parameter is rather not measurable experimentally). 
The experimental rainbow angle was determined by averaging over all collision orientations while the theoretical results reported here are for a single (Face II) orientation.

\begin{table} 
    \begin{tabular}{ c | c c }
    \hline\hline 
                                & Rainbow angle     & Impact parameter \\
                                & [degrees]         & [Bohr radius $a_0$]\\
\hline
Experimental (Average) ~\cite{Chiu_1988}   & 10.0              & -- \\ 
B3LYP                           & 6.8               & 2.7\\
PBE                             & 3.6               & 2.7\\
LDA                             & 4.3               & 2.7\\
LDA + FA                        & 16.8              & 2.7\\
LDA + ADSIC                     & 12.5              & 3.0\\
EXX                             & 13.5              & 2.9\\
END~\cite{Jacquemin_1997}       & 14.4              & 3.0 \\
ALDA + ADSIC~\cite{Gao_2014}  & 12.3              & 2.8 \\
    \hline\hline
    \end{tabular} 
    \caption{{\label{tab:rainbow} Rainbow angles and the corresponding impact parameter where it is achieved. 
    Comparison between the experimental result (the impact parameter is not accessible) and different theoretical models.
    }} 
\end{table}

\begin{figure}[h!]
\begin{center}
\includegraphics[width=1.00\columnwidth]{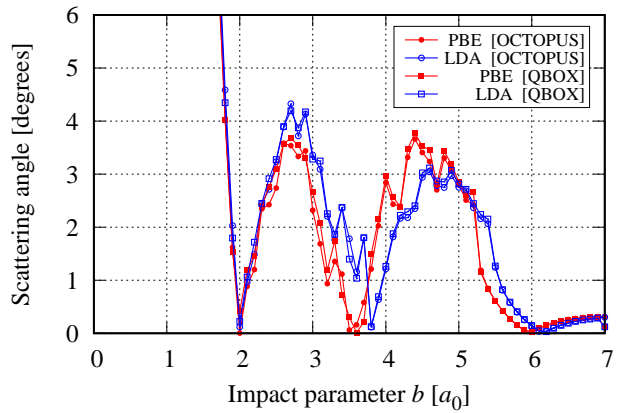}
\caption{{\label{fig:octopus_qbox} (color online). Scattering angle $\theta$ as a function of impact parameter $b$ (in units of Bohr radius $a_0$) obtained from both the real-space code (\textsc{Octopus}) and the plane-wave code (\textsc{Qbox}) for Face II orientation. 
The agreement between two independent software implementations of Ehrenfest-TDDFT shows that the rugged deflection curve is a feature of the functionals without self-interaction corrections and not an artifact due to the numerical implementation.%
}}
\end{center}
\end{figure}

Table~\ref{tab:rainbow} compares different rainbow angle values.
This comparison shows that B3LYP, PBE and LDA severely underestimate the experimental rainbow angle, showing that their results are qualitatively incorrect. 
LDA + FA, on the other hand, produces a maximum deflection angle that is $68\%$ larger than the experimental value. 
Higher-level self-interaction-free approaches like LDA + ADSIC and EXX overestimate the rainbow angle too, but are relatively closer to the experimental results. 
The ADSIC results agree with Gao et al.~\cite{Gao_2014} and the EXX compare favorably with the HF-based method of Jaquemin \emph{et al.}~\cite{Jacquemin_1997}.
Both LDA and PBE show the same features in \textsc{Octopus} and \textsc{Qbox} as shown in Fig.~\ref{fig:octopus_qbox}.

\begin{figure}[h!]
\begin{center}
\includegraphics[width=1.00\columnwidth]{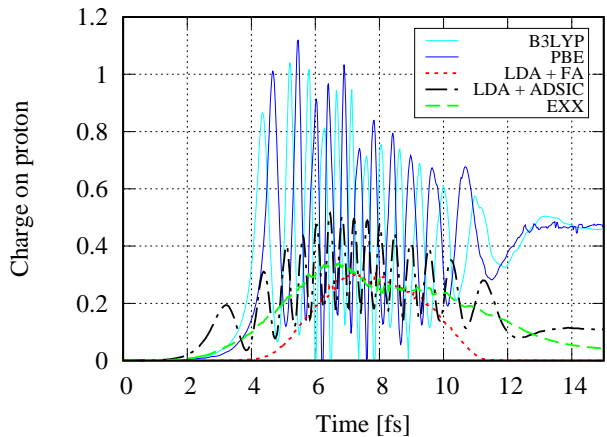}
\caption{{\label{fig:charge_on_proton_octopus} (color online). Evolution of Bader charge on the incoming proton 
as a function of simulation time for impact parameter \(b = 3.0~a_0\) (real-space code).
}}
\end{center}
\end{figure}

To understand the nature of the repulsive regime in functionals without self-interaction corrections, we plot in Fig.~\ref{fig:charge_on_proton_octopus} the charge on the proton as a function of time for $b = 3.0~a_0$. 
We define the charge on the proton using Bader charge analysis~\cite{Bader_Analysis,Bader_1991,Yu_2011}. 

The Bader analysis assigns charge to an atom, and consists in dividing space by curved closed surfaces defined by saddle points of the charge density (a minimum perpendicular to the surface). 
It is intuitively justified and practical in many molecular systems. 
Although its application to time-dependent problem is arguable, we find the overall results to be robust when compared to geometric Voronoi partitioning~\cite{Tang_2009}. 
Contrary to the Mulliken population analysis~\cite{Mulliken_1955}, Bader's is based on the electronic charge density alone~\cite{Henkelman_2006}.

\begin{figure}[h!]
\begin{center}
\includegraphics[width=1.00\columnwidth]{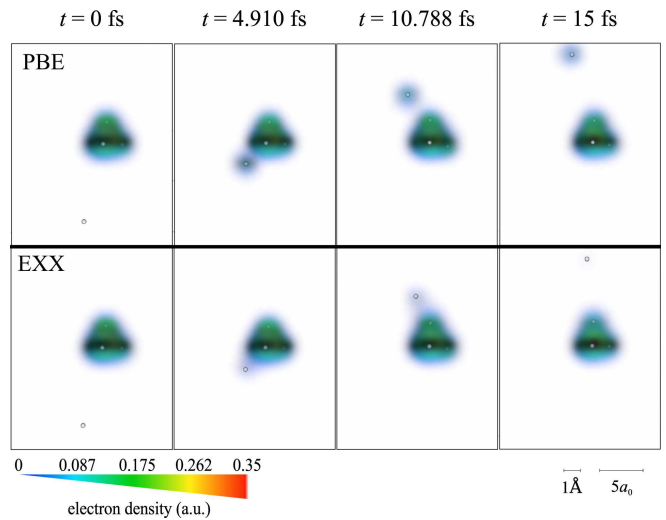}
\caption{{\label{fig:PBE_charge} (color online). Snapshots of electronic charge density distribution of $\mathrm{H^+ + CH_4}$ collision as a function of time at $b = 3.0~a_0$ 
for PBE and EXX (real-space code). 
The single gray ball represents the incoming $\mathrm{H^+}$ and the middle balls with green-contour (charge density) represent the $\mathrm{CH_4}$ atoms. 
$\mathrm{H^+}$ interacts with the $\mathrm{CH_4}$ charge density where some electron density is acquired during its trajectory and finally leaves $\mathrm{CH_4}$ with some fraction of charges (see Table~\ref{tab:final_charges}).
Orientation is the same as in Fig.~\ref{fig:dia_ch4}.
Plots are produced using the \textsc{VisIt} visulation tool~\protect\cite{HPV:VisIt}.
}}
\end{center}
\end{figure}

For B3LYP and PBE, the charge $q$ on the proton strongly oscillates, reaching a maximum value around $q = 1$. 
LDA + FA and EXX functionals have a very different behavior, there is a much smaller transfer of charge, with a maximum value around $q = 0.31$. 
In the case of LDA + ADSIC, there are oscillations in the charge, but they are smaller in amplitude and maximum value with respect to functionals without self-interaction corrections, which is rather not a surprising fact. 
The effect of the self-interaction is to make it easier for the electron to escape from the $\mathrm{CH_4}$ molecule, as the effective potential it sees is neutral; while the, more correct, self-interaction free effective potential corresponds to a system with charge $-1$, which is more attractive for the proton. 

Fig.~\ref{fig:PBE_charge} shows snapshots of the distribution of the electronic charge density as a function of time for some of the functionals (PBE and EXX). 
From these figure it is observed that there is a significant difference in the charge transferred from the $\mathrm{CH_4}$ molecule to the projectile $\mathrm{H^+}$ between functionals. 

\begin{figure}[h!]
\begin{center}
\includegraphics[width=1.00\columnwidth]{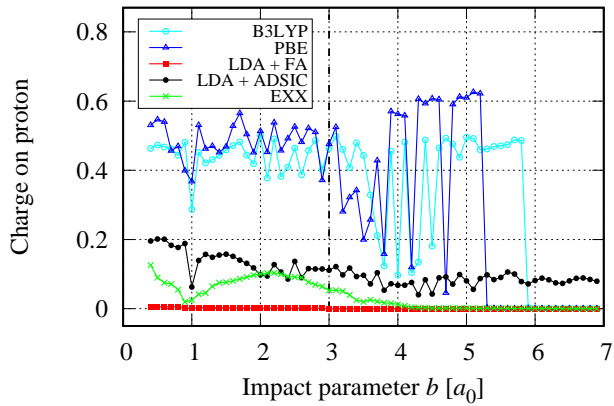}
\caption{{\label{fig:charge_impart_proton} (color online). 
Charge on the proton after the collision as a function of impact parameter $b$ (in units of Bohr radius $a_0$). All charges are measured at simulation time $t = 15~\mathrm{fs}$ using the Bader approach~\protect\cite{Bader_1991}.
The vertical line indicates the impact parameter of Fig.~\ref{fig:charge_on_proton_octopus} and Table~\ref{tab:final_charges}. For LDA + ADSIC, from $b > 8.0~a_0$ the charge on proton slowly decays to zero (not shown here).%
}}
\end{center}
\end{figure}

In Fig.~\ref{fig:charge_impart_proton}, we show the charge on the proton after the collision ($t=15~\mathrm{fs}$) for different impact parameters.
PBE and B3LYP again display uneven transfer of charge between the methane molecule and the proton at different $b$ values.
These rapid oscillations in the charge are more prominent in $3.0 \leq b \leq 6.0~a_0$ range.
For LDA + FA, the charge transfer is suppressed since electrons are bound more strongly to the molecule and all charges surrounding the proton during its trajectory returns to the molecule in the final state for all impact parameter $b$.
For the EXX case, 
some charge remains on the proton for $b < 4.0~a_0$. 
In LDA + ADSIC, the charge transfer is rather weakly dependent on impact parameter up to $\sim 8~a_0$, after which it decays smoothly to zero.

\begin{figure}[h!]
\begin{center}
\includegraphics[width=1.00\columnwidth]{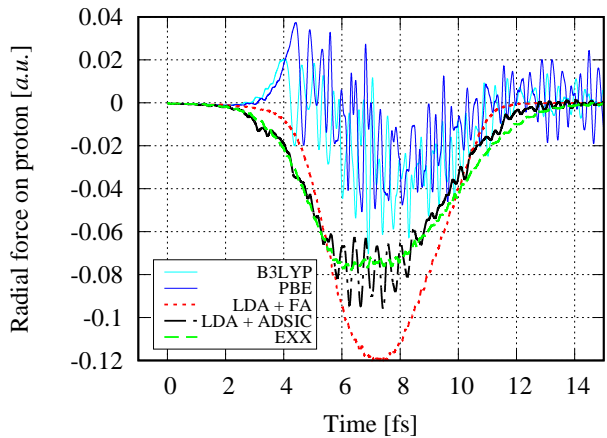}
\caption{{\label{fig:force} The radial force on proton as a function of simulation time at $b = 3.0~a_0$ (real-space code).
Radial force is taken in the direction joining the central molecule to the moving proton. 
Negative values correspond to attractive forces.
}}
\end{center}
\end{figure}

In turn, this transfer of charge produces a strong difference in the resulting force between the $\mathrm{H^+}$ and the $\mathrm{CH_4}$ molecule, and the ensuing deflection angle. 
This force is depicted in Fig.~\ref{fig:force}. 
At long ranges, 
the initial interaction is attractive; a charged particle is attracted by the dipole it induces in the molecule. 
However, when a certain amount of charge $q$ is transferred to the proton, both parts of the system are positively charged and they repel each other with a force that is proportional to $q (1-q)$. 

This repulsive force produces the positive deflection angle regime that we observe at high impact parameters (shown in Fig.~\ref{fig:scatt_octopus}) for both B3LYP and PBE. 
Moreover, since the charge in the proton oscillates, the force also does oscillate and changes sign. 
The effect of these oscillations is to reduce the overall deflection and thus yields the roughness observed in the scattering angle as a function of the impact parameter (Fig.~\ref{fig:scatt}). 

The findings of LDA + FA show, however, the opposite behavior to the functionals without self-interaction corrections; it suggests different mechanisms are playing active role in this case. 
The FA corrections add to the central region of the system an attractive potential, which makes it harder for the electron to jump to the incoming proton; it thus not only yields less charge transfer probability but also avoids the repulsive interaction. 
This attractive potential has other effects, too. 
It reduces the polarizability of the system, making the long range interaction smaller; 
for example, at time $t = 4~\mathrm{fs}$ in Fig.~\ref{fig:force} the force in LDA + FA is about half of that due to LDA + ADSIC or EXX. 
For shorter ranges, however, LDA + FA strongly attracts the proton to the molecule, showing much larger forces than other functionals. 
This strong interaction explains the overestimation compared to the experimental rainbow angle.

The functionals with self-interaction corrections exhibit radial forces that appear to be symmetric with respect to times before and after the closest approach (Fig.~\ref{fig:force}). 
This is more evident for LDA + FA, which is the functional that prevents the electron transfer the most and therefore makes the collision more elastic and, in particular, more reversible.

\begin{table} 
    \begin{tabular}{ c | c c }
    \hline\hline
                                & Charge on the proton\\
                                & [e] \\
\hline
B3LYP                           & 0.4585\\
PBE                             & 0.4668\\
LDA + FA                        & 0.0003\\
LDA + ADSIC                     & 0.1088\\
EXX                             & 0.0414\\ 
    \hline\hline
    \end{tabular}
    \caption{{\label{tab:final_charges} Final charge on the proton after the collision event ($t = 15~\mathrm{fs}$) for Face II orientation at $b = 3.0~a_0$. Charges were obtained using the Bader charge analysis~\cite{Bader_Analysis,Bader_1991,Wang_2014}.}}
    \end{table}

Another result of our simulation is the remaining charge on the proton after the collision event, shown in Table~\ref{tab:final_charges}. 
While one would expect the final charge on the proton to be an integer, our simulations yield fractional charges~\cite{Madjet_2016}. 
This is rather not a limitation of the XC functional but a property of Ehrenfest dynamics. 
Ehresfest is averaging over several potential energy surfaces that have different final charge state for the proton \cite{Tully_1990}. 
Therefore, we expect the final charge to reflect the probability distribution of the proton charge state that would be experimentally measured.
%

It is possible that the observed trends are general to other types of collision and energy scales; 
for example, as described in Ref.~\cite{Kirchner_1998} where the collision of bare ions and noble-gas atoms at $10~\mathrm{keV/amu}$ yielded fundamentally different cross-sections between LDA and the optimized potential method (OPM).

\section{Conclusion}
This paper reports our investigation on $\mathrm{H^+ + CH_4}$ collision process at $E_\text{lab} = 30~\mathrm{eV}$ at a single orientation. 
Two separate code packages \textsc{Qbox} and \textsc{Octopus}, which are independent Ehrenfest-TDDFT implementations, are used in this study for comparison and also to confirm the effects induced by self-interaction effects in the electron and ion dynamics.
The LDA, PBE and B3LYP functionals tend to produce trajectories sensitive to initial conditions and yield rugged deflection functions due to high frequency electronic oscillations. 
The inclusion of self-interaction corrections (SIC), even employing simple approximations like FA, yields smooth trajectories, 
which are, however, more comparable with previous HF-based results \cite{Jacquemin_1997}. 
The important common feature is that HF can be considered self-interaction free according to several definitions, which seems to be the most salient feature of the approximations needed to model this regime of collisions.

The different XC functionals studied in this investigation, can be grouped conveniently into SIC and non-SIC and they are very sensitive to collision dynamics. 
The signatures of these approximations are explicitly visible in all our reported findings -- scattering angle, forces and charges.

It is not hard to imagine that the spurious effects we have seen due to self-interaction are not a feature exclusive of the $\mathrm{H^+ + CH_4}$ collision. 
We expect that similar effects would be observed in other systems, especially closed shell targets. 
This allows us to theorize that in all cases that involve charge-transfer collisions, it is necessary to use self-interaction-free functionals or to include an ad-hoc correction term. 
This is necessary to correctly describe the binding of the outermost electron, and avoid too much, or too little, charge to move to the ion, changing the nature of the ion-molecule interaction.
Fortunately, simple approximations, like ADSIC, improve considerably the results of self-interacting potentials with a negligible increment in the computational cost. 
We believe that this property allows Ehrenfenst-TDDFT to offer a good trade-off between accuracy and computational cost for the simulation of molecular collisions, in particular for large systems that are not accessible to more accurate theories.
%
%
%

\section{Acknowledgments}
Authors are thankful to Professor C.~A. Weatherford, Dr. E.~R. Schwegler, Dr. G. Avenda{\~{n}}o-Franco and Dr.~K.~J.~Reed for many useful discussions. This was a joint project between the Department of Physics, Florida A\&M University and Lawrence Livermore National Laboratory. This work was performed under the auspices of the U.S. Department of Energy by Lawrence Livermore National Laboratory under Contract DE-AC52-07NA27344. Computing support for this work came from the Lawrence Livermore National Laboratory Institutional Computing Grand Challenge program. E. E. Quashie and B. C. Saha thankfully acknowledge the support by National Nuclear Security Administration (NNSA), Award Number(s) DE-NA0002630.

\bibliographystyle{apsrev4-1}
\bibliography{bibliography/converted_to_latex.bib%
}

\end{document}